\documentclass{PoS} 

\usepackage{latexsym}
\usepackage{graphicx}
\usepackage{bm}

\def\LorInd{{\mu_1\mu_2\cdot\cdot\cdot\mu_n}}
\def\LorIndtwo{{\mu_2\cdot\cdot\cdot\mu_n}}
\def\Dcc{\stackrel{\,\leftrightarrow}{D}}

\title{Nucleon structure in lattice QCD with dynamical domain-wall fermions quarks} 

\ShortTitle{Nucleon structure with dynamical DWF} 

\author{Huey-Wen Lin\\ 
      Thomas Jefferson National Accelerator Facility, 12000 Jefferson Avenue, Newport News, VA 23606\\ 
      E-mail: \email{hwlin@jlab.org}
}
\author{\speaker{Shigemi Ohta}\\ 
      Institute of Particle and Nuclear Studies, KEK, Tsukuba, Ibaraki 305-0801, Japan\\
      RIKEN-BNL Research Center, Brookhaven National Laboratory, Upton, NY 11973\\
      Physics Department, Sokendai Grad.\ Univ.\ Advanced Studies, Tsukuba, Ibaraki 305-0801, Japan\\
      E-mail: \email{shigemi.ohta@kek.jp} 
} 
\author{RBC and RBC/UKQCD Collaborations}

\abstract{
We report RBC and RBC/UKQCD lattice QCD numerical calculations of nucleon electroweak matrix elements with dynamical domain-wall fermions (DWF) quarks.
The first, RBC, set of dynamical DWF ensembles employs two degenerate flavors of DWF quarks and the DBW2 gauge action.
Three sea quark mass values of 0.04, 0.03 and 0.02 in lattice units are used with about 200 gauge configurations each.
The lattice cutoff is \(a^{-1}\sim1.7 {\rm GeV}\) and the spatial volume is about \(({\rm 1.9 fm})^3\).
Despite the small volume, the ratio of the isovector vector and axial charges \(g_{_A}/g_{_V}\) and that of structure function moments \(\langle x \rangle_{u-d}/\langle x \rangle_{\Delta u - \Delta d}\) are in agreement with experiment, and show only very mild quark mass dependence.
The second, RBC/UK, set of ensembles employs one strange and two degenerate (up and down) dynamical DWF quarks and Iwasaki gauge action.
The strange quark mass is set at 0.04, and three up/down mass values of 0.03, 0.02 and 0.01 in lattice units are used.
The lattice cutoff is \(a^{-1}\sim1.6 {\rm GeV}\) and the spatial volume is about \(({\rm 3.0 fm})^3\).
Even with preliminary statistics of 25-30 gauge configurations, the ratios \(g_{_A}/g_{_V}\) and \(\langle x \rangle_{u-d}/\langle x \rangle_{\Delta u - \Delta d}\) are consistent with experiment and show only very mild quark mass dependence.
Another structure function moment, \(d_1\), though yet to be renormalized, appears small in both sets.

\vspace{-205mm}\parbox{\textwidth}{\flushright\large\rm \hfill KEK-TH-1107, RBRC-614}\vspace{205mm}} 

\FullConference{XXIVth International Symposium on Lattice Field Theory\\ 
                July 23-28, 2006\\ 
                Tucson, Arizona, USA} 

\begin{document} 

\section{Nucleon structure on the lattice}

Nucleon isovector vector and axial charges, \(g_{_V}\) and \(g_{_A}\), are defined in neutron \(\beta\) decay form factors: \(g_{_V} = \lim_{q^2\rightarrow 0} g_{_V}(q^2)\) with
\[
\langle n| V^-_\mu(x) | p \rangle
= i\bar{u}_n [\gamma_\mu g_{_V}(q^2)
             +q_\lambda \sigma_{\lambda\mu} g_{_M}(q^2) ] u_p e^{-iqx},
\]
and \(g_{_A} = \lim_{q^2\rightarrow 0} g_{_A}(q^2)\) with
\[
\langle n| A^-_\mu(x) | p \rangle
= i\bar{u}_n \gamma_5
             [\gamma_\mu g_{_A}(q^2)
             +q_\mu g_{_P}(q^2) ] u_p e^{-iqx}.
\]
Their ratio, \(g_{_A}/g_{_V}\), is very accurately measured as 1.2695(29) \cite{PDBook}.

These form factors are calculable on the lattice, but quite often prohibitively complicated if one uses staggered or Wilson fermions.
The staggered fermions, with their weird flavor/taste structure, make even definition of nucleon impractical.
The Wilson fermions make the necessary current renormalization complicated due to explicit violation of chiral symmetry.
The domain-wall fermions (DWF)  \cite{Kaplan:1992bt,Shamir:1993zy,Furman:1995ky,Blum:2000kn}, with their exponential suppression of chiral symmetry breaking, make such renormalizations easy.
In particular the vector and axial currents should share the same renormalization.
Thus the ratio \(g_{_A}/g_{_V}\) is naturally renormalized in DWF lattice calculations \cite{Sasaki:2003jh}.

Structure functions are known from lepton deep inelastic scattering off nucleon \cite{Taylor:1991ew}, the cross section of which is factorized in terms of the leptonic and hadronic tensors.
The hadronic tensor, \(W_{\mu\nu}\), is decomposed into symmetric unpolarized and antisymmetric polarized parts:
\begin{eqnarray}
W^{\{\mu\nu\}}(x,Q^2) &=&  \left( -g^{\mu\nu} + \frac{q^\mu q^\nu}{q^2}\right)
 {F_1(x,Q^2)} \nonumber + 
\left(P^\mu-\frac{\nu}{q^2}q^\mu\right)\left(P^\nu-\frac{\nu}{q^2}q^\nu\right)
\frac{F_2(x,Q^2)}{\nu} \nonumber\\
W^{[\mu\nu]}(x,Q^2) &=& i\epsilon^{\mu\nu\rho\sigma} q_\rho
\left(\frac{S_\sigma}{\nu}({g_1(x,Q^2)}+
                           {g_2(x,Q^2)}) - 
\frac{q\cdot S P_\sigma}{\nu^2}{g_2(x,Q^2)} \right)\nonumber
\end{eqnarray}
with $\nu = q\cdot P$, $S^2 = -M^2$, $x=Q^2/2\nu$.
The unpolarized structure functions are $F_1(x,Q^2)$, $F_2(x,Q^2)$, and the polarized are $g_1(x,Q^2)$, $g_2(x,Q^2)$.
Their moments are described in terms of Wilson's operator product expansion:
\begin{eqnarray}
2 \int_0^1 dx x^{n-1} {F_1(x,Q^2)} 
&=& \sum_{q=u,d} c^{(q)}_{1,n}(\mu^2/Q^2,g(\mu))\: \langle x^n \rangle_{q}(\mu)
+{{\cal O}(1/Q^2)},
\nonumber \\ 
\int_0^1 dx x^{n-2} {F_2(x,Q^2)} 
&=& \sum_{f=u,d} c^{(q)}_{2,n}(\mu^2/Q^2,g(\mu))\: \langle x^n \rangle_{q}(\mu)
+{{\cal O}(1/Q^2)}, 
\nonumber \\
2\int_0^1 dx x^n {g_1(x,Q^2)} 
  &=& \sum_{q=u,d} e^{(q)}_{1,n}(\mu^2/Q^2,g(\mu))\: \langle x^n \rangle_{\Delta q}(\mu)
+{{\cal O}(1/Q^2)},  \nonumber
 \\
2\int_0^1 dx x^n {g_2(x,Q^2)}
  &=& \frac{1}{2}\frac{n}{n+1} \sum_{q=u,d} [e^{q}_{2,n}(\mu^2/Q^2,g(\mu))\: d_n^{q}(\mu) \nonumber\\
  &&-  2 e^{q}_{1,n}(\mu^2/Q^2,g(\mu))\: \langle x^n \rangle_{\Delta q}(\mu)] + 
{{\cal O}(1/Q^2)},\nonumber
\label{in1}
\end{eqnarray}
where $c_1$, $c_2$, $e_1$, and $e_2$ are the perturbatively known Wilson coefficients and ${\langle x^n \rangle_{q}(\mu)}$, ${\langle x^n \rangle_{\Delta q}(\mu)}$ and $d_n(\mu)$ are calculable on the lattice as forward nucleon matrix elements of certain local operators.
Again the conventional staggered or Wilson fermions complicate such lattice calculations for the same respective reasons as discussed about the form factors.
The DWF calculations are simpler because of easier renormalizations due to good chiral symmetry.
In particular the first moments \(\langle x \rangle_{u-d}\) (quark momentum fraction) and \(\langle x \rangle_{\Delta u - \Delta d}\) (quark helicity fraction) share a common renormalization and so their ratio is naturally renormalized in DWF calculations \cite{Orginos:2005uy}.

In this report we discuss dynamical DWF lattice QCD calculations of nucleon electroweak form factor ratio, \(g_{_A}/g_{_V}\), the structure function ratio, \(\langle x \rangle_{u-d}/\langle x \rangle_{\Delta u - \Delta d}\), and a polarized structure function moment \(d_1\).  The former two are naturally renormalized but \(d_1\) is yet to be renormalized.

\section{RBC and RBC/UK dynamical DWF ensembles}

The numerical calculations are performed with two different sets with dynamical domain-wall fermions (DWF) quarks.
One employs two degenerate flavors of DWF quarks.  The other employs three flavors, one strange and two degenerate and lighter up and down quarks.

In generating the former set we used the rectangular-improved DBW2 gauge action \cite{Takaishi:1996xj,deForcrand:1999bi}.
The gauge coupling is set at 0.8 and the lattice cut off turned out to be about 1.7 GeV \cite{Aoki:2004ht}.
The four-dimensional lattice size is \(16^3\times 32\), corresponding to about \(({\rm 1.9 fm})^3\) spatial box. 
This set consists of three ensembles, each with different quark mass, 0.04, 0.03 and 0.02 in lattice units.  These values roughly correspond to 1, 3/4 and 1/2 of physical strange quark mass.
The pion mass for these dynamical quark mass values are about 700, 610 and 490 MeV, respectively.
The nucleon mass are about 1.5, 1.4 and 1.3 GeV.
The fifth-dimensional extent of the lattice is set as \(L_s=12\) and the domain-wall height \(M_5=1.8\).
The residual mass was measured as \(m_{\rm res}=0.00137(5)\) or about 2.5 MeV.
In this report we use  220 gauge configurations at the heavier two dynamical mass values and 175 at the lightest.
These ensembles were generated using the QCDSP computers at RIKEN-BNL Research Center and Columbia University.

The latter set was generated using another rectangular gauge action, the Iwasaki action \cite{Iwasaki:1983i}.
The gauge coupling is set at 2.13 and the lattice cutoff is about 1.6 GeV \cite{Meifeng_Lat_2006,Tweedie_Lat_2006}.
There are two different  four dimensional lattice sizes, \(16^3\times 32\) and \(24^3\times 64\), that corresponds to about \(({\rm 2.0 fm})^3\) and \(({\rm 3.0 fm})^3\) spatial box respectively.
In this report we concentrate on the latter, larger volume.
The strange quark mass is set at about the physical value, 0.04 in lattice units.
Three ensembles are generated with degenerate up and down quark mass set at 0.03, 0.02 and 0.01 respectively.
The pion mass for these dynamical quark mass values are about 620, 520 and 390 MeV, the nucleon mass about 1.4, 1.3 and 1.2 GeV.
The fifth-dimensional extent of the lattice is \(L_s=16\) and the domain-wall height \(M_5=1.8\).
The residual mass was measured as \(m_{\rm res}=0.00308(3)\) or about 4.8 MeV.
In this report we use 25-30 gauge configurations at each up/down mass value.
These ensembles were generated using the QCDOC computers at RIKEN-BNL Research Center, Columbia University and Edinburgh University.

\section{Lattice nucleon matrix elements}

Our lattice formulation follows the standard.  The nucleon two-point function is defined as 
\(
G_{_N}(t) = {\rm Tr}[
(1+\gamma_t) \sum_{\vec{x}} \langle TB_1(x)B_1(0) \rangle]
\)
using \(B_1 = \epsilon_{abc} (u_a^T C \gamma_5d_b)u_c\) for proton.
The form-factor three-point functions are defined as
\[
G_{_V}^{u,d}(t,t') = {\rm Tr} [
(1+\gamma_t) \sum_{\vec{x}'}\sum_{\vec{x}}
\langle TB_1(x') V_t^{u,d}(x) B_1(0) \rangle ]
\]
and
\[
G_{_A}^{u,d}(t,t') = \frac{1}{3}\sum_{i=x,y,z} {\rm Tr} [
(1+\gamma_t)\gamma_i\gamma_5 \sum_{\vec{x}'}\sum_{\vec{x}}
\langle TB_1(x') A_i^{u,d}(x) B_1(0) \rangle ]
\]
with fixed \(t'=t_{\rm source}-t_{\rm sink}\) set at about 1.5 fm in physical unit and varying \(t\) smaller than \(t'\).
The ratio of their isovector combinations,
\(\displaystyle
\left[\frac{G_{_A}^u(t,t')-G_{_A}^d(t,t')}{G_{_V}^u(t,t')-G_{_V}^d(t,t')}
\right]^{\rm lattice},
\)
directly yields the renormalized isovector charge ratio \((g_{_A}/g_{_V})^{\rm ren}\).

Unpolarized structure funtion moments are obtained from three-point functions,
$$
 \frac{1}{2} \sum_s \langle P,S|{{\cal O}^{q}_{\{\LorInd\}}}
 |P,S\rangle =
 2 {\langle x^{n-1}\rangle_q}(\mu) [ P_{\mu_1}P_{\mu_2}
\cdot\cdot\cdot P_{\mu_n}+
\cdot\cdot\cdot -({\rm trace})]\nonumber
$$
$$
{{\cal O}^{q}_{\LorInd}} = \bar{q}
\left[ \left(\frac{i}{2}\right)^{n-1}
\gamma_{\mu_1} \Dcc_{\mu_2}\cdot\cdot\cdot\Dcc_{\mu_n} - ({\rm trace}) \right] q
$$
On the lattice we can measure $\langle x\rangle_q$, $\langle x^2\rangle_q$ and $\langle x^3\rangle_q$.
Higher moment operators mix with lower dimensional ones:
operators belonging in irreducible representations of $O(4)$ transform
reducibly under the lattice Hyper-cubic group.
Only $\langle x\rangle_q$ can be measured with $ \vec P = 0$.

Polarized structure function moments are obtained by
$$
 -\langle P,S|{\cal O}^{5q}_{\{\sigma\LorInd\}} |P,S\rangle=
 \frac{2}{n+1}\langle x^{n}\rangle_{\Delta q}(\mu)
 [ S_\sigma P_{\mu_1}P_{\mu_2}\cdot\cdot\cdot P_{\mu_n}+\cdot\cdot\cdot
 -({\rm traces})]
$$
$$
{\cal O}^{5q}_{\sigma\LorInd} = \bar{q}
\left[ \left(\frac{i}{2}\right)^{n}\gamma_5 \gamma_{\sigma} \Dcc_{\mu_1}
\cdot\cdot\cdot \Dcc_{\mu_n} - ({\rm traces}) \right] q
$$
$$
 \langle P,S|
{\cal O}^{[5]q}_{[\sigma\{\mu_1]\LorIndtwo\}} 
 |P,S\rangle=
 \frac{1}{n+1}d_n^{ q}(\mu)
 [ (S_\sigma P_{\mu_1} - S_{\mu_1} P_{\sigma})P_{\mu_2}
\cdot\cdot\cdot P_{\mu_n}+\cdot\cdot\cdot -({\rm traces})]
$$
$$
{\cal O}^{[5]q}_{[\sigma\mu_1]\LorIndtwo} = \bar{q}
\left[\left(\frac{i}{2}\right)^{n}\gamma_5 \gamma_{[\sigma} \Dcc_{\mu_1]}\cdot\cdot\cdot \Dcc_{\mu_n} - ({\rm traces}) \right] q.
$$
On the lattice we can measure $\langle 1 \rangle_{\Delta q}$ ($g_A$), $\langle x\rangle_{\Delta q}$, $\langle x^2\rangle_{\Delta q}$, $d_1$, and $d_2$. 
Only $\langle 1 \rangle_{\Delta q}$, $\langle x\rangle_{\Delta q}$, and $ d_1$ can be measured with  $ \vec P = 0$.

The lattice bare values of these quantities need be renormalized before comparison with experiments.
Such renormalizations are being performed non-perturbatively within Rome/Southampton RI/MOM scheme \cite{Martinelli:1995ty,Blum:2001sr} and will be reported soon.
In this report the two naturally renormalized ratios, \(g_{_A}/g_{_V}\) and \(\langle x \rangle_{u-d}/\langle x \rangle_{\Delta u-\Delta d}\), and unrenormalized \(d_1\) are discussed.

\section{Numerical results}

\begin{figure}[t]
\includegraphics[width=0.49\textwidth]{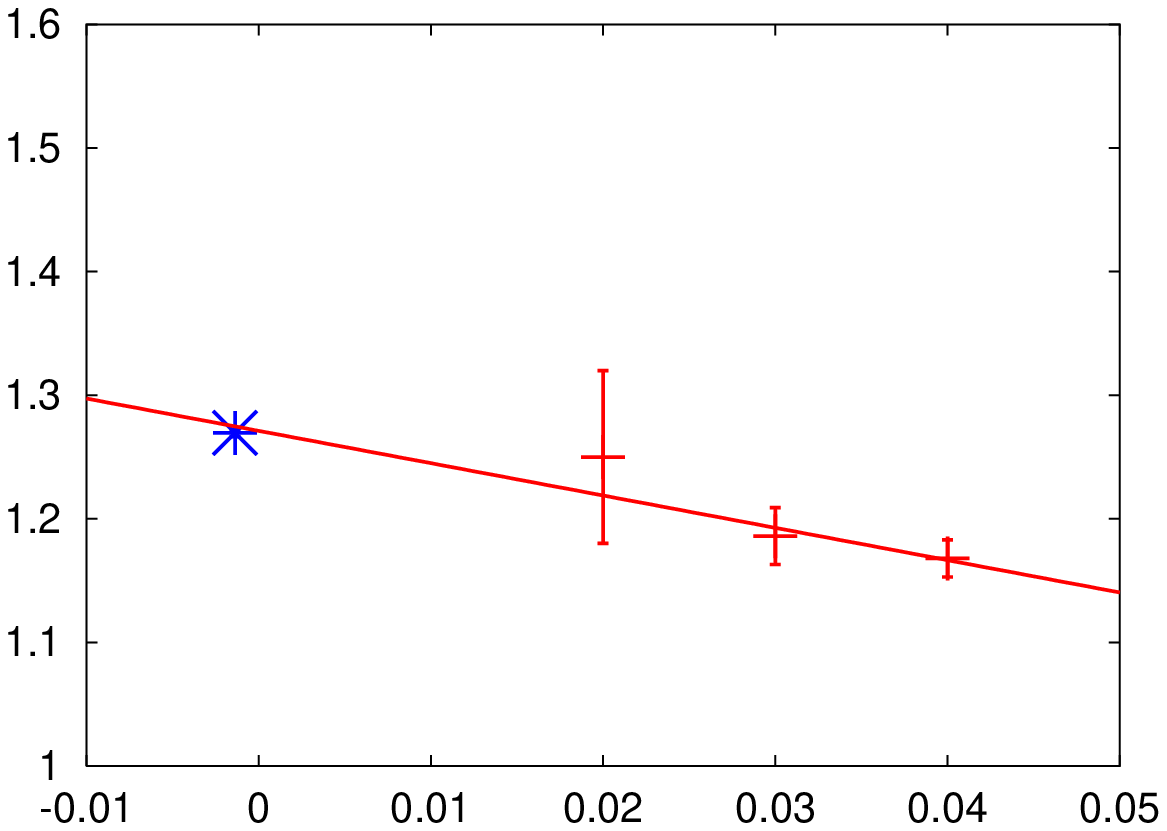}
\includegraphics[width=0.49\textwidth]{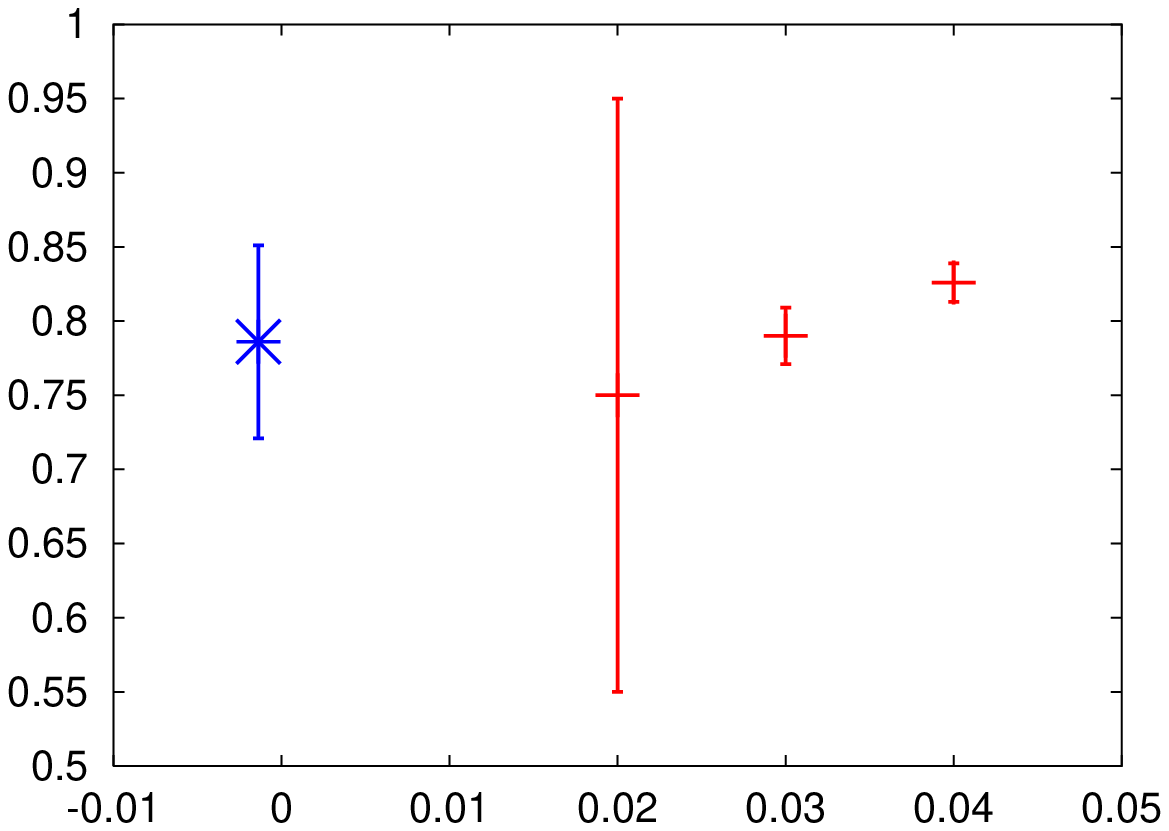}
\caption{
\label{fig:2f}
Two-flavor dynamical DWF results (red cross) for renormalized ratios, \(g_{_A}/g_{_V}\) (left) and \(\langle x \rangle_{u-d}/\langle x \rangle_{\Delta u-\Delta d}\) (right) plotted against the bare quark mass.  Respective experimental values are given as blue burst at \(m_{ud}=-m_{\rm res} = -0.00137\).
Linear fit to \(g_{_A}/g_{_V}\) is also presented.}
\end{figure}
Let us first discuss the results obtained from the RBC two-flavor dynamical DWF ensembles.
In figure \ref{fig:2f} we plot the renormalized ratios, \(g_{_A}/g_{_V}\) (left) and \(\langle x \rangle_{u-d}/\langle x \rangle_{\Delta u-\Delta d}\) (right) against the bare up/down quark mass in lattice unit.
Both show only mild quark mass dependence.
\(g_{_A}/g_{_V}\) appears in good agreement with the experiment.
Linear fit to calculated values yields a value of 1.27(5) in the chiral limit, in comparison with the experimental value of 1.2695(29).
\(\langle x \rangle_{u-d}/\langle x \rangle_{\Delta u-\Delta d}\) also appears in broad agreement with the experiment.

\begin{figure}[b]
\includegraphics[width=0.49\textwidth]{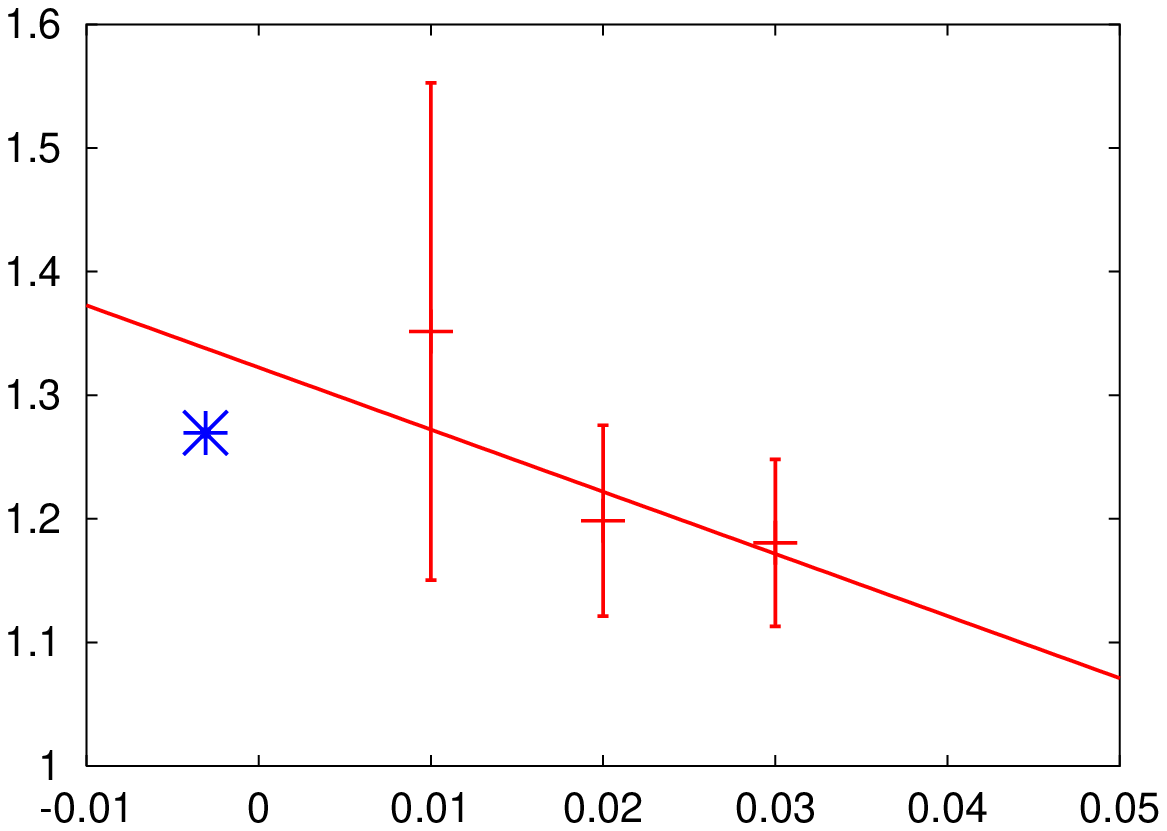}
\includegraphics[width=0.49\textwidth]{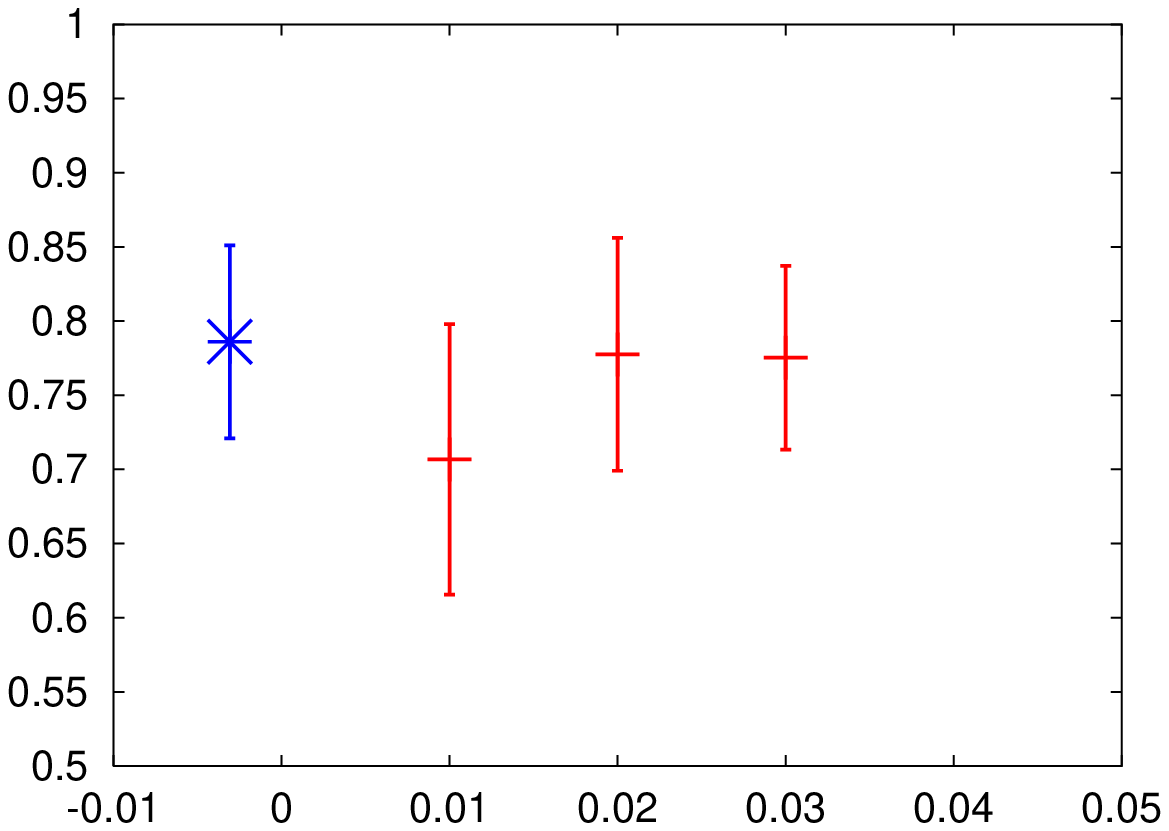}
\caption{
\label{fig:3f}
Three-flavor dynamical DWF results (red cross) for renormalized ratios, \(g_{_A}/g_{_V}\) (left) and \(\langle x \rangle_{u-d}/\langle x \rangle_{\Delta u-\Delta d}\) (right) plotted against the bare up/down quark mass.  Respective experimental values are given as blue burst at \(m_{ud}=-m_{\rm res} = -0.00308\).
Linear fit to \(g_{_A}/g_{_V}\) is also presented.}
\end{figure}
Now we turn to the results obtained from the RBC/UKQCD three-flavor dynamical DWF ensembles.  These results are still preliminary with statistics of 25-30 configurations.
In figure \ref{fig:3f} we plot the renormalized ratios, \(g_{_A}/g_{_V}\) (left) and \(\langle x \rangle_{u-d}/\langle x \rangle_{\Delta u-\Delta d}\) (right) against the bare up/down quark mass in lattice unit.
Again both show only mild quark mass dependence and appear in broad agreement with respective experiments.
Linear fit to calculated \(g_{_A}/g_{_V}\) yields a value of 1.32(11).

\begin{figure}
\includegraphics[width=0.49\textwidth]{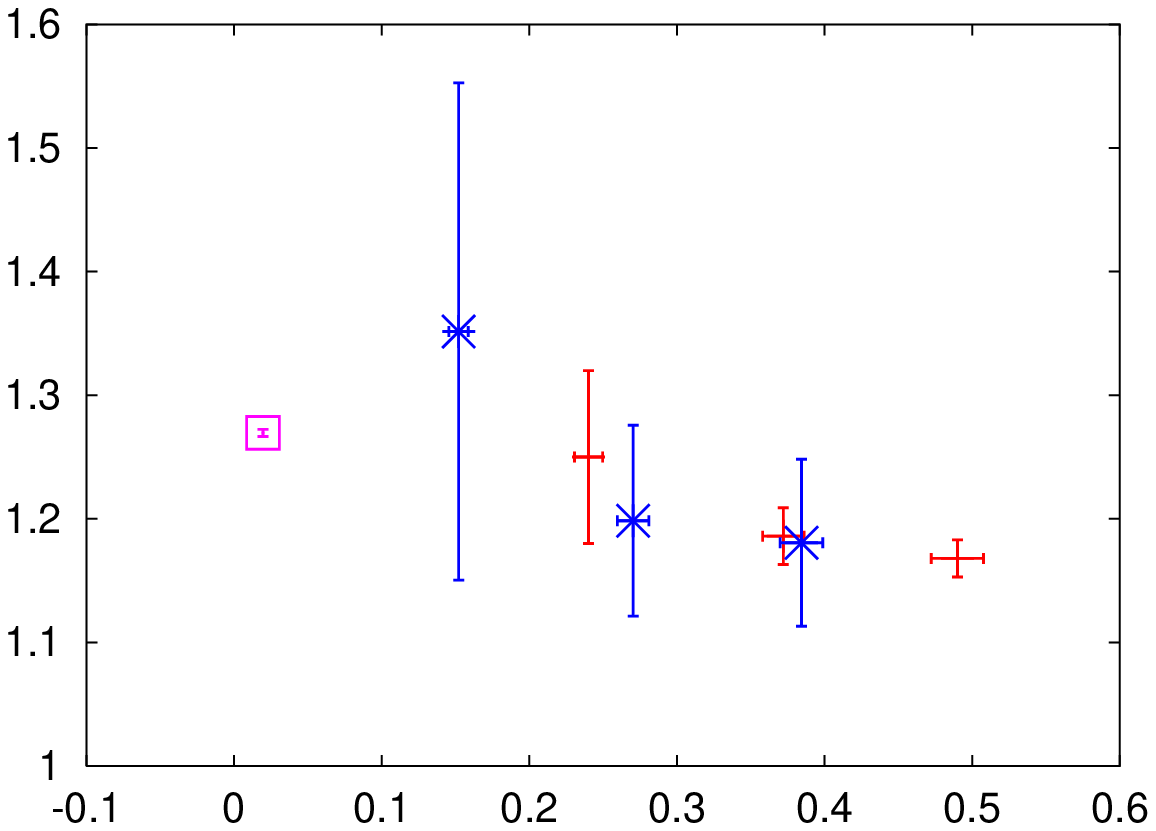}
\includegraphics[width=0.49\textwidth]{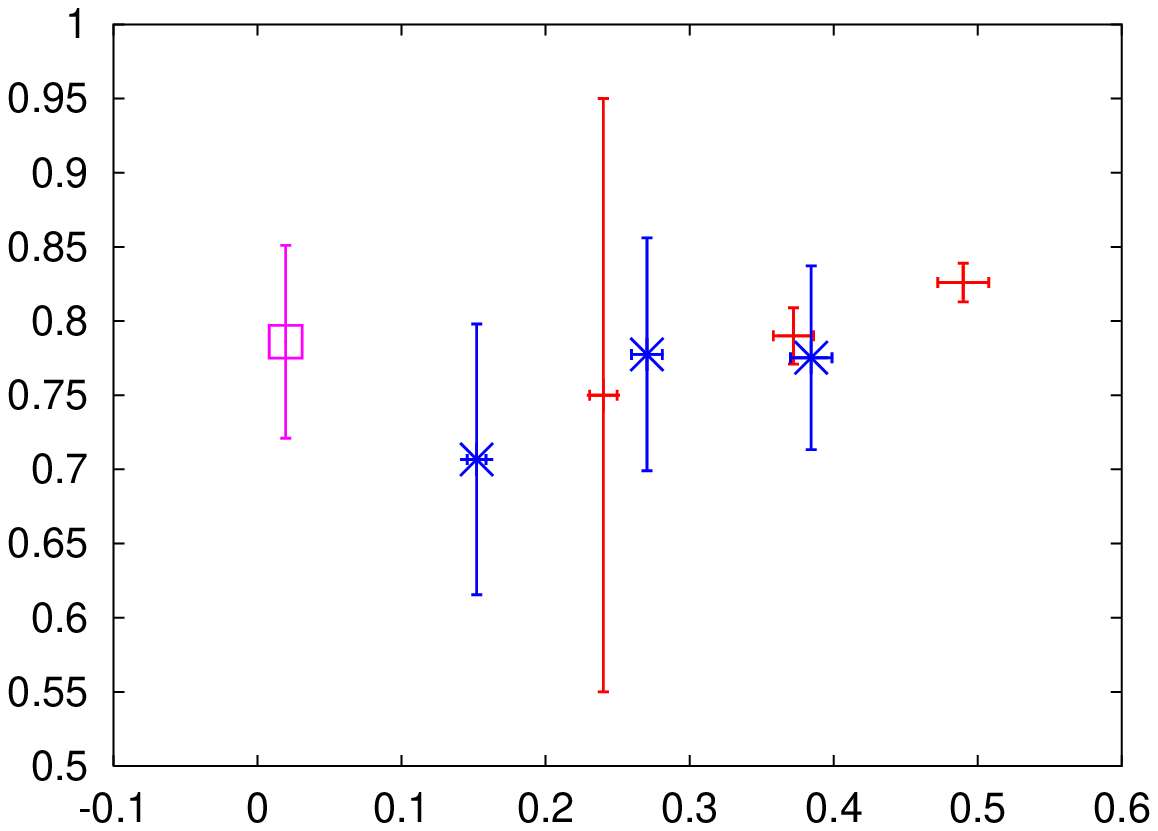}
\caption{
\label{fig:mpi2}
Renormalized ratios, \(g_{_A}/g_{_V}\) (left) and \(\langle x \rangle_{u-d}/\langle x \rangle_{\Delta u-\Delta d}\) (right) plotted against \(m_\pi^2\) in \({\rm GeV}^2\).  Two- (\(+\)) and three-flavor (\(\times\)) calculations and experiment (\(\Box\)).}
\end{figure}

Since these calculated ratios are naturally renormalized, they can be compared with each other (see figure \ref{fig:mpi2}) in physical scale such as pion mass squared, \(m_\pi^2\).  They appear mutually consistent and appear in broad agreement with the experiments.

\begin{figure}[b]
\includegraphics[width=0.49\textwidth]{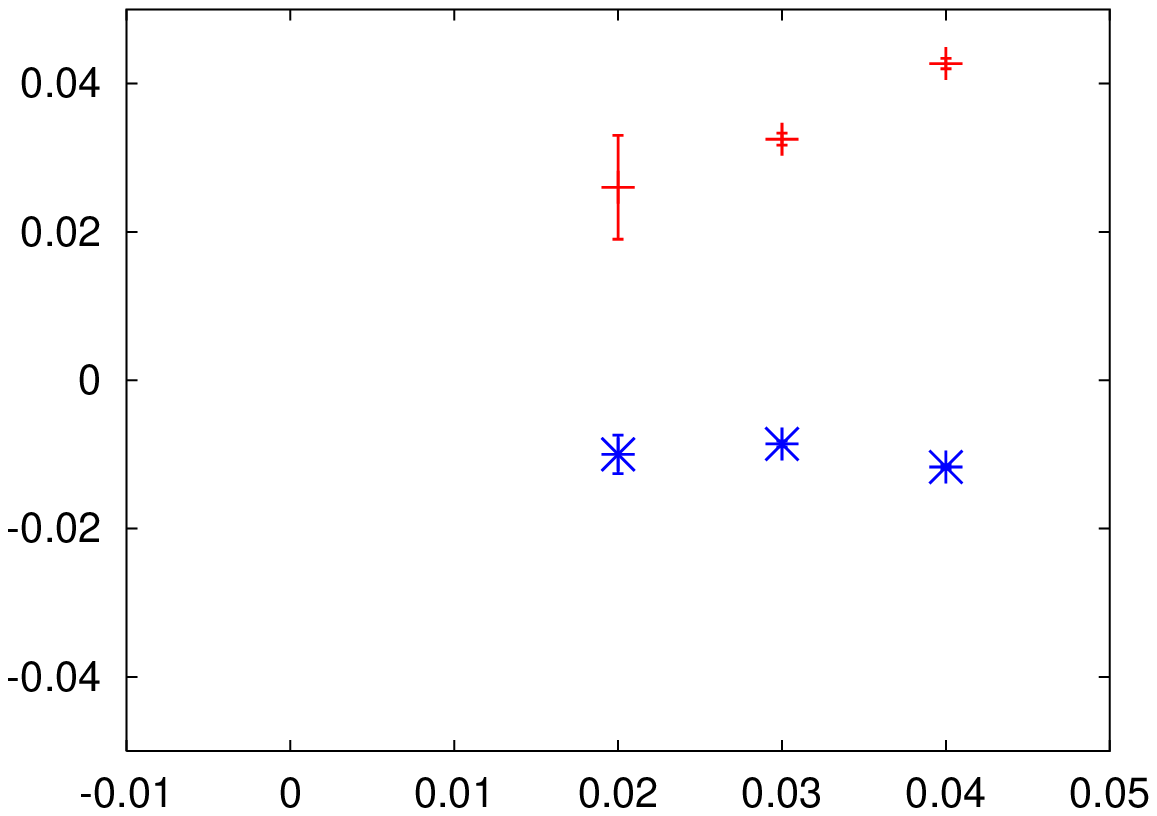}
\includegraphics[width=0.49\textwidth]{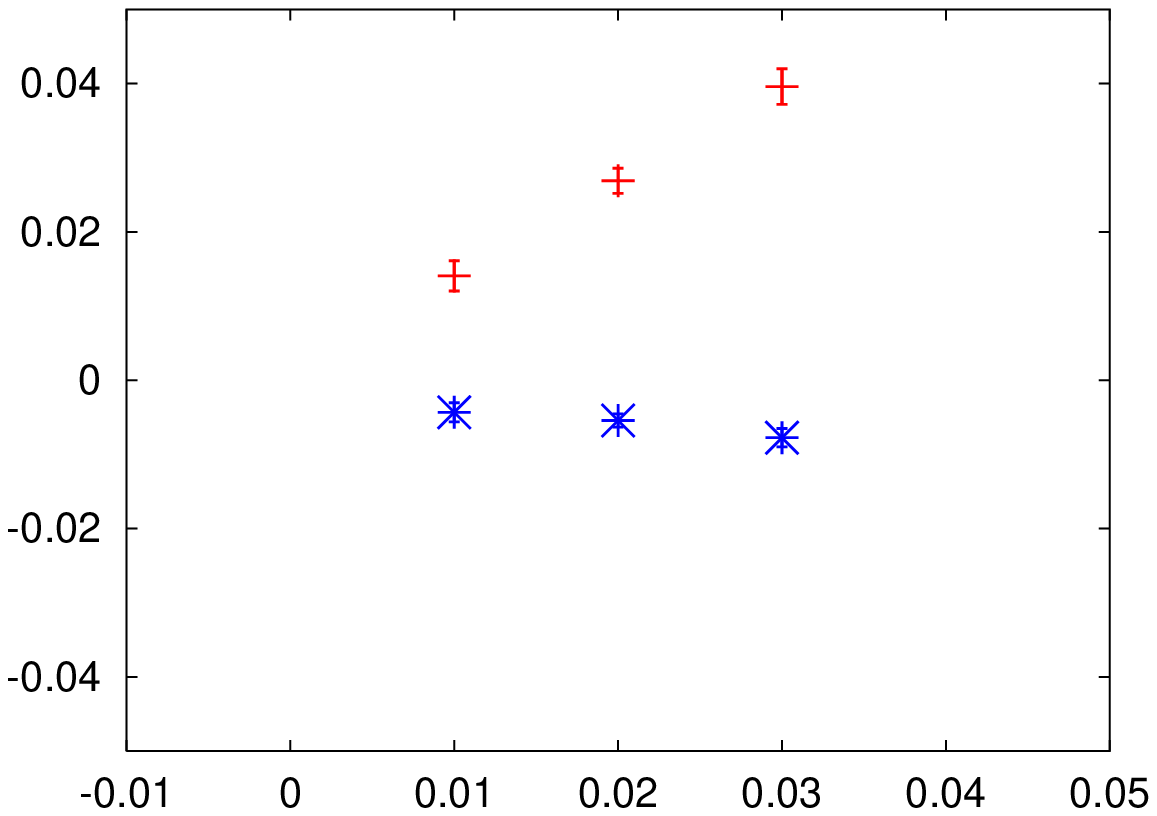}
\caption{
\label{fig:d1}
Up (\(+\)) and down (\(\times\)) contributions to bare polarized structure function moment \(d_1\), with two (left) and three (right) dynamical flavors, against the bare up/down quark mass.  They appear small.}
\end{figure}

In figure \ref{fig:d1} we present the up and down contributions to the polarized structure function moment \(d_1\), with two (left) and three (right) dynamical flavors.
This quantity summarizes the twist-3 part of the \(g_2\) polarized structure function.
Wandzura-Wilzcek relation, \(\displaystyle g_2(x) = - g_1(x) + \int_x^1 \frac{dy}{y} g_1(y)\), dictates it should be small \cite{Wandzura:1977qf}.
However it need not be small in confining theories \cite{Jaffe:1991qh}.
Our result, though yet to be renormalized, suggests it is small.

\section{Conclusions}

RBC two-flavor DBW2+DWF dynamical calculations are almost complete:
the lattice cutoff is about 1.7 GeV, and the residual mass is \(m_{\rm res}\) = 0.00137(5) in lattice units.
With three ensembles at degenerate up/down quark mass of \(m_{\rm sea}\) = 0.04, 0.03, and 0.02,
the renormalized ratios, \(g_{_A}/g_{_V}\) and \(\langle x\rangle_{u-d}/\langle x\rangle_{\Delta u-\Delta d}\), appear in agreement with experiment, despite rather small volume.
They show only mild quark mass dependence, and the linear chiral extrapolation for the former yields a value \(g_{_A}/g_{_V}\) = 1.27(5).
Non-perturbative renormalizations for individual quantities are on the way.
In particular, a polarized structure function moment, \(d_1\), appears small, though not renormalized.

RBC/UKQCD three-flavor, Iwasaki+DWF dynamical calculations are ongoing:
the lattice cutoff is about 1.6 GeV and the residual mass is \(m_{\rm res}\) = 0.00308(3).
With the strange quark mass of 0.04, three ensembles are being generated at up/down quark mass of 0.03, 0.02, and 0.01.
Results from the large-volume (3-fm across) ensembles, though preliminary at small statistics, are encouraging.
The renormalized ratios, \(g_{_A}/g_{_V}\) and \(\langle x\rangle_{u-d}/\langle x\rangle_{\Delta u-\Delta d}\), consistent with experiment.
Again they show only mild quark mass dependence, and linear fit yields a value \(g_{_A}/g_{_V}\) = 1.32(11).
\(d_1\) appears small again, though not renormalized.

We thank the members of RBC and UKQCD Collaborations, especially Tom Blum, Kostas Orginos, and Shoichi Sasaki.
The two-flavor dynamical DWF ensembles were generated using the QCDSP computers at RIKEN-BNL Research Center and Columbia University.
We Thank RIKEN, Brookhaven National Laboratory and to the U.S.\ Department of Energy for providing the facilities essential for the completion of this work.
The three-flavor dynamical DWF ensembles were generated using the QCDOC computers at RIKEN-BNL Research Center, Columbia University and Edinburgh University.
We thank Peter Boyle, Dong Chen, Mike Clark, Norman Christ, 
Saul Cohen, Calin Cristian, Zhihua Dong, Alan Gara, Andrew Jackson, 
Balint Joo, Chulwoo Jung, Richard Kenway, Changhoan Kim, 
Ludmila Levkova, Xiaodong Liao, Guofeng Liu, Robert Mawhinney, 
Shigemi Ohta, Konstantin Petrov, Tilo Wettig and Azusa Yamaguchi 
for developing the QCDOC machine and its software.
This development  and the resulting QCDOC equipment used in this work were funded by the U.S.\ DOE grant DE-FG02-92ER40699, PPARC JIF grant PPA/J/S/1998/00756 and by RIKEN.



\end{document}